\documentclass[prl,twocolumn,preprintnumbers,amsmath,amssymb]{revtex4}

\usepackage{graphicx}

\begin{document}
\title{ \bf Quantum phase transitions in bilayer quantum Hall systems }
\author{\bf  Jinwu Ye and Longhua Jiang  }
\affiliation{ Department of Physics, The Pennsylvania State
University, University Park, PA, 16802 }
\date{\today}

\begin{abstract}
   We construct a quantum
   Ginsburg-Landau theory to study the quantum phase transition from the excitonic superfluid (ESF) to
   a possible pseudo-spin density wave (PSDW) at some intermediate distances
   driven by the magneto-roton minimum collapsing at a finite wavevector.
   We analyze the properties of the PSDW
   and explicitly show that a square lattice is the favorite lattice. We suggest that correlated
   hopping of vacancies in the active and passive layers in the PSDW state leads
   to very large and temperature dependent drag consistent with the experimental data.
   Comparisons with previous microscopic numerical calculations are
   made. Further experimental implications are given.
\end{abstract}
\maketitle

   {\bf Introduction. }
  Spin-polarized Bilayer Quantum Hall systems at total filling factor
  $ \nu_{T} =1 $ have been under enormous experimental and theoretical investigations over the last decade
  \cite{gold,rev}.  When the interlayer separation $ d $ is sufficiently large, the bilayer
  system decouples into two separate compressible $ \nu=1/2 $ Fermi Liquid (FL) layers.
  However, when $ d $ is smaller than a critical distance, even in the absence
  of interlayer tunneling, the system undergoes a quantum phase transition into
  a novel spontaneous interlayer coherent incompressible phase which is an excitonic  superfluid state (ESF)
  in the pseudospin channel \cite{gold,rev,fer,wen}.
   Although there are very little dissipations
   in both the ESF and FL, the experiment \cite{drag} discovered strong enhancement of drag and dissipations
   in an intermediate distance regime.
   One of the outstanding problems in BLQH is to understand
   novel phases and quantum phase transitions as the distance
   between the two layers is changed.
   If the experimental observations indicate that there is an intermediate phase between the two
   phases remains controversial. Even it exists, different scenarios are proposed for the nature of the
   intermediate phase \cite{wigner,first,stern,simon}.
   Using Hartree-Fock (HF) in the Lowest Landau Level (LLL)  or trial wavefunctions
    approximations, many authors \cite{wigner} proposed different kinds of translational symmetry
    breaking ground states as candidates of the intermediate state.
     In this paper, we develop a quantum Ginsburg-Landau (QGL) theory \cite{cbtwo}
     to study the nature of the intermediate phase.
     We propose there are two critical distances $ d_{c1} < d_{c2} $ and three phases as the distance increases.
     When $ 0 < d < d_{c1} $, the system is in the phase ordered ESF state which breaks the internal $ U(1) $ symmetry,
     when $ d_{c1} < d < d_{c2} $, the system is in a pseudo-spin density wave ( PSDW ) state
     which breaks the translational symmetry, there is a first order transition
     at $ d_{c1} $ driven by magneto-roton minimum collapsing at a finite wavevector in the pseudo-spin
     channel. When $ d_{c2} < d < \infty $, the system becomes two weakly coupled
     $ \nu =1/2 $ Composite Fermion Fermi Liquid ( FL) state.
     There is also a first order transition at $ d= d_{c2} $.
     However, disorders could smear the two first order transitions into two second order transitions.
     We construct a quantum Ginzburg-Landau theory to describe the ESF to the PSDW which break the two completely
     different symmetries and analyze in detail the properties of the
     PSDW. By using the QGL, we explicitly show that a square lattice is the favorite lattice. The correlated hopping
     of vacancies in the active and passive layers in the PSDW state leads
     to very large and temperature dependent drag consistent with the experimental data in \cite{drag}.
     Recently, the effects of small imbalance above the PSDW were studied in
     \cite{imb} and found to explain the experimental observations nicely in \cite{imbexp}.

{\bf Formalism in phase representation. }  Consider a bi-layer
system with $ N_{1} $ ( $ N_{2} $ ) electrons in top ( bottom )
layer and
   with interlayer distance $ d $ in the presence of magnetic field $ \vec{B} = \nabla \times \vec{A} $:
\begin{eqnarray}
   H & = & H_{0} + H_{int}    \nonumber  \\
   H_{0} & = &  \int d^{2} x c^{\dagger}_{\alpha}(\vec{x})
   \frac{ (-i \hbar \vec{\nabla} + \frac{e}{c} \vec{A}(\vec{x}) )^{2} }
      {2 m }  c_{\alpha}(\vec{x})                \nonumber   \\
    H_{int} & = &  \frac{1}{2} \int d^{2} x d^{2} x^{\prime} \delta \rho_{\alpha} (\vec{x} )
               V_{\alpha \beta} (\vec{x}-\vec{x}^{\prime} )  \delta \rho_{\beta } ( \vec{x^{\prime}} )
\label{first}
\end{eqnarray}
  where electrons have {\em bare} mass $ m $ and carry charge $ - e $, $ c_{\alpha}, \alpha=1,2 $ are
  electron  operators in top and bottom layers, $ \delta \rho_{\alpha}(\vec{x}) = c^{\dagger}_{\alpha} (\vec{x})
  c_{\alpha} (\vec{x} ) - \bar{\rho}_{\alpha}, \alpha=1,2 $ are normal ordered electron densities on each layer.
  The intralayer interactions
  are $ V_{11}=V_{22}= e^{2}/\epsilon r $, while interlayer interaction is $ V_{12}=V_{21}= e^{2}/ \epsilon
  \sqrt{ r^{2}+ d^{2} } $ where $ \epsilon $ is the dielectric
  constant. For simplicity, we only discuss the balanced case in this paper.
  The effects of imbalance were discussed in detail in \cite{imb,cbtwo}.

     Performing a singular gauge transformation $
  \phi_{a}(\vec{x}) = e ^{ i \int d^{2} x^{\prime} arg(\vec{x}-\vec{x}^{\prime} )
  \rho ( \vec{x}^{\prime} ) } c_{a}( \vec{x}) $
  where $ \rho ( \vec{x} ) = c^{\dagger}_{1}( \vec{x} ) c_{1}( \vec{x} ) +
  c^{\dagger}_{2}( \vec{x} ) c_{2}( \vec{x} )  $ is the total density of the bi-layer system.
  We can transform the Hamiltonian Eqn.\ref{first} into a Lagrangian of the Composite
  Boson $ \phi_{a} $ coupled to a Chern-Simon gauge field $ a_{\mu} $ \cite{cbtwo}.
  We can write the two bosons in terms of magnitude and phase $
  \phi_{a}= \sqrt{ \bar{\rho}_{a} + \delta \rho_{a} } e^{i \theta_{a}
  } $, then after absorbing the external gauge potential $ \vec{A} $ into $
   \vec{a} $, we get the Lagrangian in the Coulomb gauge \cite{cbtwo}:
\begin{eqnarray}
  {\cal L}  &  =  & i \delta \rho_{+} ( \frac{1}{2} \partial_{\tau} \theta_{+}-  a_{0} ) +
          \frac{ \bar{\rho} }{2m} [ \frac{1}{2} \nabla \theta_{+}
          - \vec{a} ]^{2}
                      \nonumber  \\
    & +  & \frac{1}{2} \delta \rho_{+} V_{+} (\vec{q} )  \delta \rho_{+}
          - \frac{ i }{ 2 \pi} a_{0} ( \nabla \times \vec{a} )
                       \nonumber   \\
    & + & \frac{i}{2} \delta \rho_{-}  \partial_{\tau} \theta_{-} +
          \frac{ \bar{\rho} }{2m}  ( \frac{1}{2} \nabla \theta_{-} )^{2}
          + \frac{1}{2} \delta \rho_{-} V_{-} (\vec{q} )  \delta \rho_{-}
\label{main}
\end{eqnarray}
     where $ \delta \rho_{\pm} = \delta \rho_{1} \pm \delta \rho_{2},
     \theta_{\pm} = \theta_{1} \pm \theta_{2} $,
     they satisfy commutation relations
     $ [ \delta \rho_{\alpha} ( \vec{x} ), \theta_{\beta}( \vec{x}^{\prime} ) ]
         = 2 i \hbar \delta_{\alpha \beta} \delta( \vec{x}-\vec{x}^{\prime} ),
     \alpha, \beta=\pm $. $ \bar{\rho} = \bar{\rho_{1}} +
     \bar{\rho_{2}},  V_{\pm}= \frac{ V_{11} \pm V_{12} }{2} $.


    It was shown in \cite{cbtwo} that the functional form in the
    spin sector in Eqn.\ref{main}  achieved from the CB theory is
    the same as the microscopic LLL+HF approximation achieved in
    \cite{rev}. However, the spin stiffness $ \frac{ \bar{\rho}}{2m} $  and the $ V_{-}(q) $ in
    Eqn.\ref{main} should be replaced by the effective ones calculated by the LLL+HF
    approximation: $ \rho_{E} $ and $ V_{E}(q)=  a-b q+ c q^{2}
    $ where the non-analytic term is due to the long-range Coulomb interaction,
    $ a \sim d^{2}, b \sim d^{2} $, but $ c  $ remains
    a constant at small distances.
    In the ESF state \cite{cbtwo}, it is convenient to integrate out
    $ \delta \rho_{-} $ in favor of the phase field $ \theta_{-} $.
 \begin{equation}
    {\cal L}_{s} = \frac{1}{ 2 V_{E}( \vec{q} ) } ( \frac{1}{2} \partial_{\tau} \theta^{-} )^{2} +
     \rho_{E} ( \frac{1}{2} \nabla \theta_{-} )^{2}
\label{xy}
\end{equation}
    where the dispersion relation of the NGM including higher orders of momentum can be extracted:
\begin{equation}
     \omega^{2} = [2 \rho_{E}  V_{E}( \vec{q} ) ] q^{2}
     = q^{2}( a-b q+ c q^{2} )
\label{disperhigh}
\end{equation}

{\bf Instability driven by the collapsing of magneto-roton minimum.}
    As shown in \cite{qgl} in the context of possible supersolid in Helium 4, the advantage to extend the dispersion relation beyond the leading
    order is that the QGL action can even capture possible phase
    transitions between competing orders due to competing interactions on microscopic length
    scales. In the following, we study the instability of the ESF state as the distance
    increases. It can be shown that the dispersion curve
    Eqn.\ref{disperhigh} and $ V_{E}( q ) $  indeed has the shape shown in Fig.1a \cite{edge}
    and Fig.1b respectively.

\begin{figure}
\includegraphics[width=8cm]{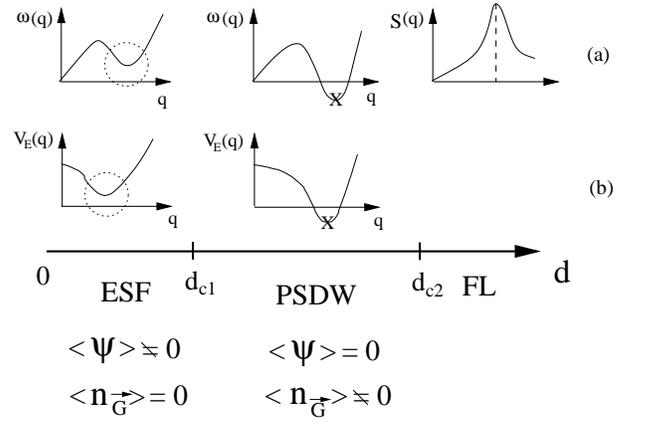}
\caption{The zero temperature phase diagram in the balanced case as
the distance between the two layers increases. ESF where $ < \psi>
\neq 0, < n_{\vec{G} } >=0 $ stands for excitonic superfluid, PSDW
where $ < \psi> = 0, < n_{\vec{G} } > \neq 0 $ stands for
pseudo-spin density wave phase, FL stands for Fermi Liquid. (a)
Energy dispersion relation $ \omega(q) $ in these phases. (b) $
V_{E}(q) $ in these phases. The cross in the PSDW means the negative
minimum value of $ V_{E}( q ) $ is replaced by the PSDW. The order
parameters are also shown. In fact, the instability happens before
the minimum touches zero.   } \label{fig1}
\end{figure}

    The phenomenon of the collapsing of the magento-roton minimum
    as the distance increases was clearly demonstrated in the numerical
    calculations in the LLL+HF approaches \cite{fer,jog} and
    detected by inelastic light scattering \cite{roton}.
    It was estimated that $ q_{0} l \sim 1 $, so the lattice constant of
    the PSDW is of the same order of magnetic length $ l $ which is $ \sim 100 \AA $. The critical
    distance $ d_{c1} $ is also of the same order of the magnetic length.
    In reality, the instability happens before the minimum touches zero.

{\bf Effective action in the dual density representation and a
Feymann relation in the pseudo-spin channel.} Because the original
instability
    comes from the density-density interaction,
    it is convenient to integrate out the phase field in favor of the
    density operator. Neglecting the vortex excitations in $ \theta_{-} $ and integrating
    out the $ \theta_{-} $ in Eqn.\ref{xy} leads to:
\begin{equation}
 {\cal L}[ \delta \rho_{-} ]=
    \frac{1}{2} \delta \rho^{-}(-\vec{q},-\omega_{n} ) [ \frac{ \omega^{2}_{n} }{ 2  \rho_{E} q^{2}}
    + V_{E} (\vec{q} ) ] \delta \rho^{-}(\vec{q},\omega_{n} )
\label{density}
\end{equation}
    where we can identify the dynamic pseudo-spin density-density correlation
    function $ S_{-}(\vec{q},\omega_{n} )
    = < \delta \rho^{-}(-\vec{q},-\omega_{n} ) \delta
    \rho^{-}(\vec{q},\omega_{n} ) > =  \frac{ 2 \rho_{E} q^{2} }{ \omega^{2}_{n} +
   v^{2}(q) q^{2} } $
    where $ v^{2}(q)= 2 \rho_{E} V_{E} (q) $ is the spin wave
    velocity.

    From the pole of the dynamic density-density correlation
    function, we can identify the speed of sound
    wave which is exactly the same as the spin wave velocity.
    This should not be too surprising. As shown in liquid $ ^{4}He $,
    the speed of sound is exactly the same as the phonon
    velocity. Here, in the context of excitonic superfluid, we
    explicitly prove that the sound speed is indeed the same as the spin
    wave velocity. From the analytical continuation
    $ i \omega_{n} \rightarrow \omega + i \delta $, we can identify the dynamic
    structure factor: $ S_{-}( \vec{q},\omega ) = S_{-}(q) \delta ( \omega
    -v(q)q ) $ where $ S_{-}(q)= \rho_{E} q \pi/v(q) $ is the equal time
    pseudo-spin correlation function shown in Fig.1. As $ q
    \rightarrow 0, S_{-}(q) \rightarrow q $.
    The {\sl Feymann relation }  in BLQH which relates the dispersion relation to the equal
    time structure factor is
\begin{equation}
     \omega (q) =  \frac{ \rho_{E} \pi  q^{2} }{  S_{-}(q) }
\end{equation}
    which takes exactly the same form as
    the Feymann relation in superfluid $ ^{4} He $.
    Obviously, the $ V_{E}(q) $ in the Fig.1b leads to the magneto-roton dispersion $ \omega^{2}= q^{2} V_{E}(q)
    $ in the Fig.1a.

     Because the instability is happening at $ q=q_{0} $ instead of at $ q=0 $, the vortex excitations in
     $ \theta_{-} $ remain {\em uncritical} through the transition. Integrating them out will not generate
     any singularities except the interactions among the pseudo-spin density $ \delta \rho_{-}
     $. Expanding $ V_{E}(q) $ near the minimum $ q_{0} $ in the Fig.
     1 leads to the quantum Ginsburg-Landau action to describe the ESF to the PSDW transition:
\begin{eqnarray}
 {\cal L}[ \delta \rho_{-} ] & = &
    \frac{1}{2} \delta \rho^{-} [ A_{\rho} \omega^{2}_{n}
    + r+ c( q^{2}-q^{2}_{0} )^{2} ] \delta \rho^{-}
        \nonumber   \\
    & + &  u ( \delta \rho_{-} )^{4} + v ( \delta \rho_{-} )^{6} + \cdots
\label{densitymin}
\end{eqnarray}
    where the momentum and frequency conservation in the quartic and
    sixth order terms is assumed, $  A_{\rho} \sim \frac{ 1 }{ 2 \rho_{E} q_{0}^{2}} $ which
    is non-critical across the transition.
    While the corresponding quantity $ A_{\theta} \sim S_{-}(q) $
    in the phase representation Eqn.\ref{xy} is divergent, so
    Eqn.\ref{xy} breaks down as $ q \rightarrow q^{-}_{0} $ and may
    not be used to describe the ESF to the PSDW transition.

    In sharp contrast to the conventional classical normal liquid (NL) to
    normal solid (NS) transition \cite{tom}, the possible cubic interaction term $ ( \delta \rho_{-})^{3} $ is
    forbidden by the $ Z_{2} $ exchange symmetry between the two layers
    $ \delta \rho_{-} \rightarrow  -\delta \rho_{-} $. Note that the $   \omega^{2}_{n} $ term in the first
    term stands for the quantum fluctuations of $ \delta \rho_{-} $ which is absent in the classical
    NL to NS transition. In the following section, we will
    show that the most favorable lattice is a square lattice.

{\bf Lattice structure of the PSDW phase.}
  In Eqn.\ref{densitymin}, $ r $ which is the gap of $ V_{E}(q) $ at
  the minimum tunes the transition from the ESF to the PSDW.
  In the ESF, $ r > 0 $ and $ < \delta \rho_{-}> = 0  $ is uniform. In the PSDW, $ r < 0 $ and
  $ < \delta \rho_{-} > = \sum_{\vec{G}} n (\vec{G} ) e^{ i \vec{G} \cdot \vec{x} }, \ n(0)=0 $
  takes a lattice structure with reciprocal lattice vectors $ \vec{G} $.
  It was shown that in the classical NL
  to NS transition, due to the cubic term, a triangular lattice is
  the favorite lattice.
  However, due to the absence of the cubic
  term in Eqn.\ref{densitymin}, in the following, we will show that the favorite lattice is the
  square lattice. At mean field level, we can ignore the $ \omega $ dependence in Eqn.\ref{densitymin}.
 Substituting $ < \delta \rho_{-} > = \sum_{\vec{G}} n (\vec{G} ) e^{ i \vec{G} \cdot \vec{x} } $ into
 Eqn.\ref{densitymin} leads to:
\begin{eqnarray}
  f_{n} & = &  \sum_{\vec{G}} \frac{1}{2}  r_{\vec{G}} | n_{\vec{G}} |^{2}
    + u  \sum_{\vec{G}} n_{\vec{G}_1}
    n_{\vec{G}_2} n_{\vec{G}_3} n_{\vec{G}_4} \delta_{ \vec{G}_1 + \vec{G}_2 +
    \vec{G}_3 + \vec{G}_4,0 }  \nonumber   \\
    & + &  v \sum_{\vec{G}} n_{\vec{G}_1}
    n_{\vec{G}_2} n_{\vec{G}_3} n_{ \vec{G}_4 } n_{ \vec{G}_5 } n_{ \vec{G}_6 } \delta_{ \vec{G}_1 + \vec{G}_2 +
    \vec{G}_3+ \vec{G}_4 + \vec{G}_5 + \vec{G}_6, 0} + \cdots
\label{compare}
\end{eqnarray}
   where $ u $ could be positive or negative, $ v > 0 $ is always
   positive to keep the system stable.

   From Eqn.\ref{compare}, we can compare the ground state energy of the two most
   commonly seen lattices: square lattice and triangular lattice.
   Square ( triangular ) lattice has $ m=4 $ ( $ m=6 $ ) shortest
   reciprocal lattice vectors $ G= 2\pi/a $ where $ a $ is the lattice constant in a given layer.
   Following \cite{tom}, one can simplify Eqn.\ref{compare} to:
\begin{equation}
  f_{\alpha} = \frac{1}{2} r_{\vec{G}} | n_{\vec{G}} |^{2}+ u_{\alpha} | n_{\vec{G}}|^{4}+ v_{\alpha} | n_{\vec{G}} |^{6}
\label{simple}
\end{equation}
   where for the quadratic and quartic terms,
   all the contributions come from the paired reciprocal lattice
   vectors. After the scaling  $ n_{\vec{G}} \rightarrow m^{-1/2} n_{\vec{G}}
   $, the quadratic term become the same for both lattices, but the quartic
   terms are still different $ u_{\alpha}=3(1-1/m)u $\cite{tom}, so $ u_{\Box}=2\frac{1}{4}u,  u_{\triangle} =2 \frac{1}{2} u $.
   For the sixth order term, in square lattice, all the contributions are still from the
   paired reciprocal lattice vectors, however, in triangular
   lattice, there is an {\em additional} contribution from a
   triangle where $ \vec{G}_1 + \vec{G}_2 + \vec{G}_3 =0 $. After
   very careful counting, we find that $ v^{p}_{\alpha}=5( 3- 9/m + 8/m^{2} )v $ from the paired contribution
   and $ v^{tri}=5/6v $ from the additional contribution from the
   triangle in a triangular lattice, so $ v_{\Box}=6 \frac{1}{4} v, v_{\triangle}=9 \frac{4}{9} v $.
  The mean field phase diagram of Eqn.\ref{simple} is well known:
  If $ u > 0 $ ( $ u < 0 $ ), there is 2nd ( 1st ) order
  transition, there is a tri-critical point at $ u=0 $. Minimizing $
  f $ with respect to $ n_{\vec{G}} $ leads to $  n^{2}_{\alpha} =
  (-2u_{\alpha} + \sqrt{ 4 u^{2}_{\alpha}-6 v_{\alpha} r } )/6v_{\alpha} $ which holds for
  both $ u_{\alpha} > 0 $ and $ u_{\alpha} < 0 $. Obviously, $ n_{\Box} \neq  n_{\triangle}$.
  From Eqn.\ref{simple}, we can see
  that if $ u > 0 $, because $ u_{\Box} < u_{\triangle}, v_{\Box} < v_{\triangle} $,
  for any given $ n $, $ f_{\Box}(n) <  f_{\triangle}(n)
  $, then $ f_{\Box}(n_{\Box} ) <  f_{\Box}( n_{\triangle} ) < f_{\triangle}(
  n_{\triangle}) $. If $ u < 0 $,  it is easy to show that
  when $ n^{2} > n^{2}_{c}= - \frac{9 u}{108 v } $, $  f_{\Box}(n) <  f_{\triangle}(n)
  $. Because in the two lattice states, $ r < u^{2}_{\alpha}/2 v_{\alpha}
  $, $ n^{2}_{\Box} > -\frac{u_{\Box}}{2 v_{\Box} }= - \frac{9u}{50v} > n^{2}_{c},
       n^{2}_{\triangle} > -\frac{u_{\triangle}}{2 v_{\triangle} }= - \frac{9u}{68v} >
       n^{2}_{c} $, so in this regime, we still have  $ f_{\Box}(n_{\Box} ) <  f_{\Box}( n_{\triangle} ) < f_{\triangle}(
  n_{\triangle}) $. So we conclude that in any case, the square lattice is the favorite
  lattice. It has three elastic constants instead of two.
  Neglecting zero-point quantum fluctuations, $ < \delta \rho_{-}
  > = \sum_{i} \delta( \vec{x} -\vec{R_{i}} )- \sum_{i} \delta( \vec{x} -\vec{R_{i}}-\vec{l}
  ) $ where the $ \vec{l}=\frac{1}{2}( \vec{a}_{1}+ \vec{a}_{2}  ) $
  is the shift of the square lattice in the bottom layer relative to that in the top layer ( Fig.2).

\begin{figure}
\includegraphics[width=6cm]{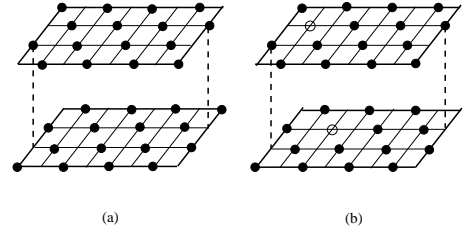}
\caption{ (a) The charge distribution of
    the  PSDW in a square lattice. The "up " pseudo-spins take
    sublattice A, while the "down" pseudo-spins take sublattice B.
    (b) In general, there are quantum and thermal fluctuation generated vacancies in both layers
    denoted by a  $ \bigcirc $.   }
\label{fig2}
\end{figure}

  This PSDW state not only breaks the translational
  symmetry, but also the $ Z_{2} $ exchange symmetry.
  It is very rare to get a 2d square lattice, because it is not a
  close packed lattice. Due to the special $ Z_{2} $ symmetry of
  BLQH, we show it indeed  can be realized in BLQH.
  The system is compressible with gapless phonon excitations
  determined by the 3 elastic constants.
  It is very interesting to see if {\em in-plane} soft X-ray or light
  scattering experiments  \cite{rev} can directly test the existence of the
  PSDW when $ d_{c1} < d < d_{c2} $.
  Note that the light scattering intensity may be diminished by a Debye-Waller factor
  due to the large zero-point quantum fluctuations in the PSDW \cite{qgl}.

{\bf  Vacancies, disorders and Coulomb Drag in the PSDW state.}
  In principle, the $ \delta \rho_{+} $
  mode in Eqn.\ref{main} should also be included.
  It stands for the translational (or sliding ) motion of the PSDW lattice.
  However, any weak disorders will pin this PSDW state. Therefore, the $ \delta \rho_{+}
  $ mode can be neglected.
  Disorders will smear the 1st order
  transition from the ESF to the PSDW into a 2nd order transition.
  It was argued in \cite{gold} that disorders in real samples are so strong that they
  may even have destroyed the ESF state,
  so they may also transfer the long
  range lattice orders of the PSDW into short range ones. This fact makes the
  observation of the lattice structure by light scattering difficult.
  The two square lattices in the top and bottom layer are locked together, so it will show huge Coulomb drag.
  However, vacancies generated by the large zero-point quantum
  fluctuations may play important roles in the drag ( Fig.2b ).
  As the distance increases to the critical
  distance $ d_{c1} $ in Fig.1, the ESF turns into the PSDW whose lattice constant $ a=\sqrt{4 \pi} l $ is
  completely fixed by the filling factor $ \nu_{1}=1/2 $, so it may not
  completely match the instability point $ 1/q_{0} $. Due to this slight mismatch, the
  resulting PSDW is very likely to be an {\em in-commensurate} solid where
  the total number of sites $ N_{s} $ may not be the same as the total number of sites $
  N $ even at $ T=0 $.
  As the distance increases further $ d_{c1} < d < d_{c2} $ in Fig.1, the PSDW
  lattice constant is still {\em locked} at $ a =\sqrt{4 \pi} l $.
  Assuming zero-point quantum fluctuations favor vacancies
  over interstitials \cite{qgl}, so there are vacancies  $ n_{0} $ even at $ T=0 $ in each layer.
  At finite temperature,
  there are also thermally generated vacancies $ n_{a}(T) \sim e^{-
  \Delta_{v}/T } $ where $ \Delta_{v} $ is the vacancy excitation
  energy. So the total number of vacancies at any $ T $ is $ n_{v}(T) =n_{0}+ n_{a}(T) $.
  Obviously, the vacancies in top layer are strongly correlated with
  those in the bottom layer. As shown in \cite{pump}, the correlated character of hopping
  transports of the vacancies between the active and
  passive layers can lead to a very large drag. We can estimate the
  resistance in the active layer as $ R \sim 1/n_{v}(T) $.
  As shown in \cite{pump}, the drag resistance in the passive layer is $ R_{D} \sim \alpha_{D} R $
  where $ \alpha_{D} $ should not be too small. This temperature
  dependence is indeed consistent with that found in the experiment
  \cite{drag}:  starting from $ 200 mK $, $ R_{D} $ increases exponentially until
  to  $ 50 mK $ and then saturates. This behavior is marked
  different than that at both  small and large  distance where the system is in the ESF and FL regimes respectively.
  These vacancies also lead to the finite Hall drag in the PSDW.
  We conclude that in the presence of disorders, all the properties of the PSDW are consistent with  the experimental
  observations in \cite{drag} on the intermediate phase.
  The effects of very small imbalance on the ESN phase was investigated in \cite{imb} and was also
  found to be consistent with the experimental data in \cite{imbexp}.

{\bf Discussions.}  We compare the results achieved from the QGL
theory with those achieved from the microscopic LLL+HF approximation
in \cite{wigner}. Especially, Cote, Brey and Macdonald
    in \cite{wigner} numerically found that the lowest energy lattice structure of the PSDW is a square
    lattice. But it is not known if the HF+LLL calculations can describe the transition from the ESF to the PSDW.
    The QGL theory presented in this paper circumvents the difficulty
    associated with the unknown wavefunction at any finite $ d $ \cite{wave} and treat both the interlayer
    and the intralayer correlations on the same footing, so
    can be used to capture competing
    orders on microscopic length scales and naturally leads to the PSDW as
    the intermediate state which breaks translational symmetry. The theory puts the ESF state and
    the PSDW state on the same footing, characterize the symmetry breaking patterns in the two
    states by corresponding order parameters and describe the universality
    class of the quantum phase transition between the two states. We explicitly showed that
    the square lattice is the most favorite lattice. There are quantum fluctuation generated vacancies in the
    PSDW which lead to the unusual temperature dependence of the Coulomb drag observed
    in the experiment \cite{drag}. The QGL theory is complementary to and goes well beyond the
    previous microscopic calculations.  Recent experiments \cite{spin} found the real spin also plays important
    roles in some BLQH samples with large tunneling gaps. Putting the spin into the theory remains an
    important open problem.



    J.Ye thank B. Halperin for helpful discussions and J. K. Jain for pointing out Ref.\cite{pump} to us.

\end{document}